\documentclass[epj,referee]{svjour}
\usepackage{amssymb}
\usepackage{amsmath}
\usepackage{graphicx}
\usepackage{dcolumn}
\usepackage{bm}

\begin{document}

\title{Preparation of a Stable and Maximally Entangled State of Two Distant
Qutrits Trapped in Separate Cavities}
\author{Chang-shui Yu\and X. X. Yi\and He-shan Song\thanks{\emph{Email address:}%
hssong@dlut.edu.cn} \and D. Mei} \institute{School of Physics and
Optoelectronic Technology, Dalian University of Technology, Dalian
116024, P. R. China}
\date{Received: date / Revised version: date}

\abstract{ We have proposed a simple scheme to entangle two distant
qutrits trapped in separate optical cavities. The quantum
information of each qutrit is skillfully encoded on the degenerate
ground states of a pair of atoms, hence the entanglement between
them is relatively stable against spontaneous emission. In
Lamb-Dicke limits, it is not necessary to require coincidence
detections, which will relax the conditions for the experimental
realization. The scheme is robust against the inefficient
detections. \PACS{
      {}{42.50.Dv}
      {}{03.67.Mn}\and
      {}{03.65.Ud}
    } }
\maketitle

Entanglement, in particular the entanglement between distant particles, is
not only a key ingredient for the tests of quantum nonlocality [1], but also
an important physical resource in achieving tasks of quantum computation and
quantum communication [2]. Hence, generation of entangled states and its
further applications are immensely important. A lot of schemes have been
proposed to generate entangled states [3-9], but most of the schemes focused
on the generation of entanglement of two qubits or more qubits. However the
entanglement between two qubits (E.g. EPR pairs) and even many qubits (For
example, GHZ or W states) can not be competent for all the tasks of quantum
information processing. In particular, higher-dimensional entanglement has
recently attracted increasing interests: A maximally entangled state of two
qudits is necessary in general, if an unknown quantum state of qudit will be
teleported exactly [10]; A known quantum state can not be remotely prepared,
unless an entangled state of two qudits has been provided [11,12]; Moreover,
cryptographic protocols based on entangled qutrits [13-16] have been shown
to be more efficient and secure than those based on qubit systems; Recently
Ref. [17] has shown that teleportation can also be implemented in faith,
even though a non-maximally entangled states defined in higher dimension can
be provided.

Based on the demand of entangled qutrits, there have been some schemes
[18-22] to generate entangled qutrits. However, these schemes are mainly
focused on the preparation of entangled qutrits at the confined location
(for example, both are trapped in a single cavity) and the generation of
entangled photons of qutrits. In particular, the partial quantum information
of a qutrit is encoded on the excited state of atoms (ions) in some schemes,
which means that the entanglement of the qutrits is fragile (not stable). So
far as we know, few schemes have been found to effectively entangle \textit{%
distant} atoms or ions of qutrits. In this paper, we propose a
simple
scheme to generate \textit{stable} maximally entangled state of two \textit{%
distant} qutrits with the help of linear optical elements. It should
first be worth noting that, although we will employ the same
atom-cavity interaction mechanism as our previous work [23], it is
not a simple extension but we address a new physical problem by a
creative design of quantum optical circuits. In our scheme, the most
important is that the quantum information of each qutrit is
skillfully encoded on the degenerate ground states of a pair of
separate atoms trapped in optical cavities, respectively, which
leads to a relatively stable entangled state. In addition, The key
of the scheme is the indistinguishability of photons emitting from
the entangled atoms, which has been widely employed to entangle
distant qubits [24-29]. Our scheme is shown to be robust against the
inefficient detections. In particular, in Lamb-Dicke limits, we do
not require the coincidence detections of photons, which can
dramatically relax the conditions of practical realization.

The systems we consider here are the same to those in Ref. [24]
where two atoms trapped in separate cavities can be robustly
entangled by simultaneously detecting the leakage photons. Here, we
will show that the expanded version can help us to entangle two
distant qutrits without
coincidence detections. We consider two pairs of identical three-level $%
\Lambda $-type atoms trapped in four separated identical one-sided optical
cavities $A_{1}$, $A_{2}$, $B_{1}$ and $B_{2}$, respectively, with cavities $%
A_{i}$ at Alice's side and cavities $B_{i}$ at Bob's. See FIG. 1
(a). Each atom has an excited state $\left\vert e\right\rangle $ and
two degenerate ground states $\left\vert g_{l}\right\rangle $ and
$\left\vert g_{r}\right\rangle $. The transitions $\left\vert
g_{l}\right\rangle \rightarrow \left\vert e\right\rangle $ and
$\left\vert g_{r}\right\rangle \rightarrow \left\vert e\right\rangle
$ are strongly coupled to left- and right- circularly polarizing
cavity modes, respectively. Our atomic level structure can be
achieved by Zeeman sublevels [33] and has been realized to entangle
two atoms [34]. The quantum information of a qutrit is encoded in
the ground states of a pair of atoms as follows.
\begin{eqnarray}
\left\vert 0\right\rangle _{\beta } &=&\left\vert g_{l}g_{l}\right\rangle
_{\beta },\left\vert 2\right\rangle _{\beta }=\left\vert
g_{r}g_{r}\right\rangle _{\beta },  \notag \\
\left\vert 1\right\rangle _{\beta } &=&\frac{1}{\sqrt{2}}\left( \left\vert
g_{l}g_{r}\right\rangle _{\beta }+\left\vert g_{r}g_{l}\right\rangle _{\beta
}\right) ,
\end{eqnarray}%
with $\beta =A,B$ denoting Alice or Bob. The experimental setup is sketched
in FIG. 1 (b). We suppose the four atoms are all initially prepared in their
excited states and cavities in the vacuum state. If all the four detectors
are clicked, Alice and Bob can share a stable and maximally entangled
bipartite quantum state of qutrits:%
\begin{equation}
\left\vert \Psi \right\rangle _{AB}=\frac{1}{\sqrt{3}}(\left\vert
0_{A}0_{B}\right\rangle +\left\vert 1_{A}1_{B}\right\rangle +\left\vert
2_{A}2_{B}\right\rangle ).
\end{equation}

In order to demonstrate our scheme in detail, let us start with the
interaction between atoms and cavities. The Hamiltonian governing the
evolution of the four atom-cavity systems can be given in the interaction
picture by (setting $\hbar =1$)%
\begin{eqnarray}
H_{j} &=&\lambda _{L}\left\vert e\right\rangle _{jj}\left\langle
g_{l}\right\vert a_{L}^{j}+\lambda _{R}\left\vert e\right\rangle
_{jj}\left\langle g_{r}\right\vert a_{R}^{j}+h.c.,  \notag \\
j &=&A_{\alpha },B_{\alpha },\alpha =1,2,
\end{eqnarray}%
where $L,$ $R$ denote the left- and right- circularly polarizing cavity
modes, $a_{k}^{j\dag },a_{k}^{j}$ are the creation and annihilation
operators of the $k$ mode in the $j$ cavity and $\lambda _{k}$ is the
coupling constant. The upper levels $\left\vert e\right\rangle $ can decay
to the two degenerate ground states $\left\vert g_{l}\right\rangle $ and $%
\left\vert g_{r}\right\rangle $ with the rates $2\gamma _{l}$ and $2\gamma
_{r}$, respectively, and every cavity has a leakage rate $2\kappa $. Hence,
the master equation describing the evolution of density operator $\rho $ is
given by
\begin{gather}
\dot{\rho}_{j}=-i(H_{eff}\rho _{j}-\rho _{j}H_{eff}^{\dag })+2\kappa
\sum_{k=L,R}a_{k}^{j}\rho _{j}a_{k}^{j\dag }  \notag \\
+2\sum_{p=l,r}\gamma _{p}\left\vert g_{p}\right\rangle _{jj}\left\langle
e\right\vert \rho _{j}\left\vert e\right\rangle _{jj}\left\langle
g_{p}\right\vert ,j=A_{\alpha },B_{\alpha },\alpha =1,2,
\end{gather}%
with
\begin{equation}
H_{eff}=H_{j}-i\kappa \sum_{k=L,R}a_{k}^{j\dag }a_{k}^{j}-i\left( \gamma
_{l}+\gamma _{r}\right) \left\vert e\right\rangle _{jj}\left\langle
e\right\vert .
\end{equation}%
So long as spontaneous emissions do not happen and cavity photons
are not leaked out, the above effective non-Hermitian Hamiltonian
(5) can be employed to describe the dynamics of the system based on
quantum jump approach [30].

\begin{figure}
\centering
\includegraphics[width=8.5cm]{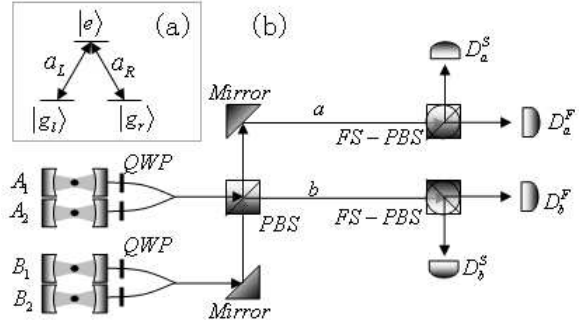}
\caption{(a) Atomic level structure. (b) Experimental setup for maximally
entangled two qutrits without coincidence detections allowed. The pair of
atoms trapped in cavities $A_{\protect\alpha },\protect\alpha =1,2$ which
are encoded by the quantum information of a qutrit are at Alice's location.
The other pair of atoms in cavities $B_{\protect\alpha },\protect\alpha =1,2$
which are encoded by another qutrit are at Bob's location. Circularly
polarizing light becomes linearly polarizing by the quarter wave plates ($QWP
$). Photons leaking out of cavities pass through the polarizing beam
splitters ($PBS$)which transmits $H$-polarizing photons and reflect $V$%
-polarizing photons. Before registered by the detectors, photons
meet the rotated $PBS$es ($FS-PBS$)with the rotation angle
$\protect\theta $ which transmit $F$-polarizing photons and reflect
$S$-polarizing photons.} \label{Fig. 1}
\end{figure}
Consider the initial state $\left\vert e\right\rangle _{j}\left\vert
0\right\rangle _{j}$, after a time $t$ the state of each system will,
respectively, become%
\begin{eqnarray}
\left\vert \psi (t)\right\rangle _{j} &=&\frac{x\left\vert e\right\rangle
_{j}\left\vert 0\right\rangle _{j}+y\left\vert g_{l}\right\rangle
_{j}\left\vert V\right\rangle _{j}+z\left\vert g_{r}\right\rangle
_{j}\left\vert H\right\rangle _{j}}{\sqrt{\left\vert x\right\vert
^{2}+\left\vert y\right\vert ^{2}+\left\vert z\right\vert ^{2}}},  \notag \\
j &=&A_{\alpha },B_{\alpha },\alpha =1,2,
\end{eqnarray}%
where
\begin{eqnarray*}
x &=&e^{-\frac{\Gamma }{2}t}\left[ \cos \left( \Omega _{k}t\right) +\frac{%
\Delta }{2\Omega _{k}}\sin \left( \Omega _{k}t\right) \right] , \\
y &=&-e^{-\frac{\Gamma }{2}t}\frac{i\sin \left( \Omega _{k}t\right) }{\Omega
_{k}}\lambda _{L}, \\
z &=&-e^{-\frac{\Gamma }{2}t}\frac{i\sin \left( \Omega _{k}t\right) }{\Omega
_{k}}\lambda _{R},
\end{eqnarray*}%
with%
\begin{eqnarray*}
\Omega _{k} &=&\sqrt{\Omega ^{2}-\Delta ^{2}/4}, \\
\Omega ^{2} &=&\lambda _{L}^{2}+\lambda _{R}^{2}, \\
\Gamma  &=&\gamma _{l}+\gamma _{r}+\kappa , \\
\Delta  &=&\kappa -\gamma _{l}-\gamma _{r}.
\end{eqnarray*}%
Here $\left\vert 0\right\rangle _{j}$ and $\left\vert V\right\rangle _{j}$, $%
\left\vert H\right\rangle _{j}$ present vacuum state and the
one-photon state with vertically and horizontally polarizing cavity
modes, respectively. From FIG. 1 (b), one can find that photons
leaking out of cavities will pass through quarter wave plates
($QWP$) which change circularly polarizing light into linearly
polarizing light. We assume left- and right- circularly polarizing
photons become vertically ($V$) and horizontally ($H$) polarizing
respectively [24]. In this sense, we have directly replaced the
circularly polarizing photons in eq. (6) by linearly polarizing
photons.

We have said that the spirit of our scheme is the
indistinguishability of photons. This usually needs the coincidence
detections of photons, otherwise the emission of a photon will lead
to a recoil of the atom [25] which destroy the indistinguishability
and lead to a failure. If our trapped atoms are restricted to
operating in the Lamb--Dicke limit, where the recoil energy does not
suffice to change the atomic motional state [25], the
indistinguishability can be preserved. Hence in Lamb-Dicke limit we
do not require the coincidence detections. Consequently we can
suppose the evolution time of every subsystem to be $\tau _{j}$. In
such an interval of time, one can obtain the state $\left\vert \psi
(\tau _{j})\right\rangle $ given by eq. (6) with the probability
\begin{equation}
P_{j}=e^{-\Gamma \tau _{j}}\left\{ \left[ \cos \left( \Omega _{k}\tau
_{j}\right) +\frac{\Delta }{2\Omega _{k}}\sin \left( \Omega _{k}\tau
_{j}\right) \right] ^{2}+\frac{\sin ^{2}\left( \Omega _{k}\tau _{j}\right) }{%
\Omega _{k}^{2}}\Omega ^{2}\right\} .
\end{equation}%
Thus the joint state of the four systems can be given by $\left\vert \Psi
\right\rangle =\otimes _{j}\left\vert \psi (\tau _{j})\right\rangle $ with
the probability $P_{1}=\prod\limits_{j}P_{j}$. From eq. (6), one can find
that the term $\left\vert e\right\rangle _{j}\left\vert 0\right\rangle _{j}$
has no contribution to the detections, hence one can safely neglect it for
simplification. As a result, one can rewrite eq. (6) at the time $\tau _{j}$
as%
\begin{eqnarray}
\left\vert \phi (\tau _{j})\right\rangle _{j} &=&\frac{1}{\Omega }\left(
\lambda _{L}\left\vert g_{l}\right\rangle _{j}\left\vert V\right\rangle
_{j}+\lambda _{R}\left\vert g_{r}\right\rangle _{j}\left\vert H\right\rangle
_{j}\right) ,  \notag \\
j &=&A_{\alpha },B_{\alpha },\alpha =1,2.
\end{eqnarray}%
In the interval of $\max_{j}\{\tau _{j}\}$, the joint state of the whole
four systems can be expressed by
\begin{equation}
\left\vert \Phi \right\rangle =\otimes _{j}\frac{1}{\Omega }\left( \lambda
_{L}\left\vert g_{l}\right\rangle _{j}\left\vert V\right\rangle _{j}+\lambda
_{R}\left\vert g_{r}\right\rangle _{j}\left\vert H\right\rangle _{j}\right) ,
\end{equation}%
with the total probability
\begin{equation}
P_{2}=\prod\limits_{j}\left[ \frac{e^{-\Gamma \tau _{j}}\sin ^{2}\left(
\Omega _{k}\tau _{j}\right) \Omega ^{2}}{\Omega _{k}^{2}}\right] .
\end{equation}

Next we will show that the state $\left\vert \Phi \right\rangle $ can
collapse to our desired state $\left\vert \Psi \right\rangle _{AB}$ given by
eq. (2) in terms of non-coincidence detections in Lamb-Dicke limits. We
assume that the four detectors are clicked at the different times which just
correspond to a permutation of the evolution time $\{\tau _{j}\}$, where we
neglect the transmission time through the linear optical elements. Because
the photons leaking out of the cavities are indistinguishable in Lamb-Dicke
limits, one can not tell which atom the photon emits from. In other words,
one can not evaluate by which atom each click is led to. Therefore the
evolution time $\{\tau _{j}\}$ is also indistinguishable for each atom. Let $%
t_{1}\leq t_{2}\leq t_{3}\leq t_{4}$ be the time sequence at which the
detectors are clicked, and which corresponds to some permutation of $\{\tau
_{j}\}$. Without loss of the generality, we suppose the detections are
clicked in turn as $D_{a}^{F}\rightarrow D_{b}^{F}\rightarrow
D_{a}^{S}\rightarrow D_{b}^{S}$ corresponding to the above time sequence.
Following FIG. 1 (b). A photon leaking out of the cavity will first pass
through a polarizing beam splitter ($PBS$) which transmits $H$- polarizing
light and reflects $V$-polarizing light and then meet a rotated $PBS$ ($%
FS-PBS$) which transforms $V$- and $H$-polarizing light as $\left\vert
V\right\rangle \rightarrow \cos \theta \left\vert F\right\rangle +\sin
\theta \left\vert S\right\rangle $ and $\left\vert H\right\rangle
\rightarrow -\cos \theta \left\vert S\right\rangle +\sin \theta \left\vert
F\right\rangle $ where $\theta $ is the rotated angle. $FS-PBS$ always
transmits $F$-polarizing light and reflects $S$-polarizing light. Therefore
if the detector $D_{a}^{F} $ is clicked at time $t_{1}$, the joint state
will collapse to%
\begin{gather}
\left\vert \Phi \right\rangle _{1}=\frac{\cos \theta \lambda _{L}}{\Omega }%
\left( \left\vert g_{l}\right\rangle _{A_{1}}\left\vert \phi \right\rangle
_{A_{2}}+\left\vert \phi \right\rangle _{A_{1}}\left\vert g_{l}\right\rangle
_{A_{2}}\right) \left\vert \phi \right\rangle _{B_{1}}\left\vert \phi
\right\rangle _{B_{2}}  \notag \\
+\frac{\sin \theta \lambda _{R}}{\Omega }\left\vert \phi \right\rangle
_{A_{1}}\left\vert \phi \right\rangle _{A_{2}}\left( \left\vert
g_{r}\right\rangle _{B_{1}}\left\vert \phi \right\rangle _{B_{2}}+\left\vert
\phi \right\rangle _{B_{1}}\left\vert g_{r}\right\rangle _{B_{2}}\right) .
\end{gather}%
When detector $D_{b}^{F}$ is clicked at time $t_{2}$, $\left\vert \Phi
\right\rangle _{1}$ collapses to
\begin{gather}
\left\vert \Phi \right\rangle _{2}=\frac{1}{\Omega ^{2}}\left[ \sqrt{2}\cos
\theta \sin \theta \lambda _{L}\lambda _{R}\left\vert 1\right\rangle
_{A}\left\vert \phi \right\rangle _{B_{1}}\left\vert \phi \right\rangle
_{B_{2}}\right. +\cos ^{2}\theta \lambda _{L}^{2}  \notag \\
\times \left( \left\vert g_{l}\right\rangle _{A_{1}}\left\vert \phi
\right\rangle _{A_{2}}+\left\vert \phi \right\rangle _{A_{1}}\left\vert
g_{l}\right\rangle _{A_{2}}\right) \left( \left\vert g_{l}\right\rangle
_{B_{1}}\left\vert \phi \right\rangle _{B_{2}}+\left\vert \phi \right\rangle
_{B_{1}}\left\vert g_{l}\right\rangle _{B_{2}}\right)  \notag \\
+\sin ^{2}\theta \lambda _{R}^{2}\left( \left\vert g_{r}\right\rangle
_{A_{1}}\left\vert \phi \right\rangle _{A_{2}}+\left\vert \phi \right\rangle
_{A_{1}}\left\vert g_{r}\right\rangle _{A_{2}}\right) \left( \left\vert
g_{r}\right\rangle _{B_{1}}\left\vert \phi \right\rangle _{B_{2}}\right.
\notag \\
+\left. \left\vert \phi \right\rangle _{B_{1}}\left\vert g_{r}\right\rangle
_{B_{2}}\right) \left. +\sqrt{2}\sin \theta \cos \theta \lambda _{R}\lambda
_{L}\left\vert \phi \right\rangle _{A_{1}}\left\vert \phi \right\rangle
_{A_{2}}\left\vert 1\right\rangle _{B}\right] .
\end{gather}%
Analogously, if at time $t_{3}$ the detector $D_{a}^{S}$ is clicked, $%
\left\vert \Phi \right\rangle _{2}$ will collapse to $\left\vert \Phi
\right\rangle _{3}$ as%
\begin{gather}
\left\vert \Phi \right\rangle _{3}=\frac{1}{\Omega ^{3}}\left[ 2\cos
^{2}\theta \sin \theta \lambda _{L}^{3}\left\vert 0\right\rangle _{A}\left(
\left\vert g_{l}\right\rangle _{B_{1}}\left\vert \phi \right\rangle
_{B_{2}}+\left\vert \phi \right\rangle _{B_{1}}\left\vert g_{l}\right\rangle
_{B_{2}}\right) \right.  \notag \\
-\sqrt{2}\lambda _{L}\lambda _{R}^{2}\sin \theta \cos \left( 2\theta \right)
\left\vert 1\right\rangle _{A}\left( \left\vert g_{r}\right\rangle
_{B_{1}}\left\vert \phi \right\rangle _{B_{2}}+\left\vert \phi \right\rangle
_{B_{1}}\left\vert g_{r}\right\rangle _{B_{2}}\right)  \notag \\
-\sqrt{2}\lambda _{R}\lambda _{L}^{2}\cos \theta \cos \left( 2\theta \right)
\left( \left\vert g_{l}\right\rangle _{A_{1}}\left\vert \phi \right\rangle
_{A_{2}}+\left\vert \phi \right\rangle _{A_{1}}\left\vert g_{l}\right\rangle
_{A_{2}}\right) \left\vert 1\right\rangle _{B}  \notag \\
-2\sin ^{2}\theta \cos \theta \lambda _{R}^{3}\left( \left\vert
g_{r}\right\rangle _{A_{1}}\left\vert \phi \right\rangle _{A_{2}}+\left\vert
\phi \right\rangle _{A_{1}}\left\vert g_{r}\right\rangle _{A_{2}}\right)
\left\vert 2\right\rangle _{B}.
\end{gather}%
\begin{figure}
\centering
\includegraphics[width=9cm,height=7.5cm]{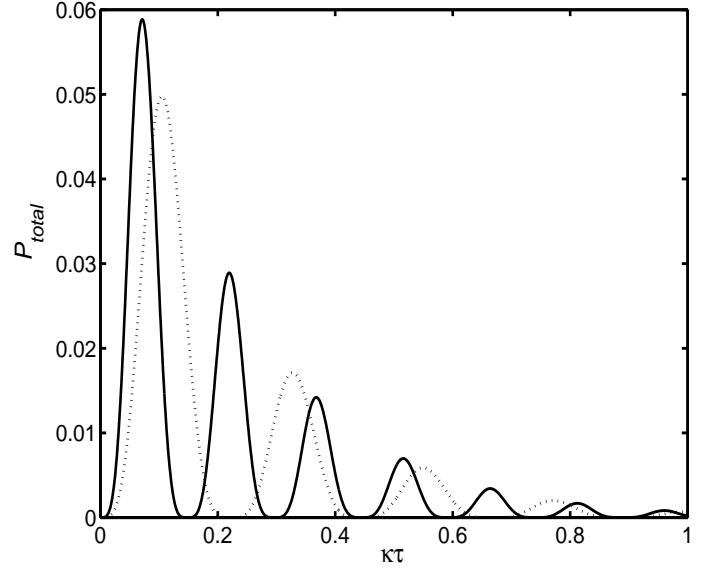}
\caption{(Dimensionless)The probability $P_{total}$ of getting a maximally
entangled state of qutrits vs $\protect\kappa \protect\tau $ with different
coupling constants. We choose $\protect\tau _{j}=\protect\tau $, $\protect%
\gamma _{l}=\protect\gamma _{r}=0.1\protect\kappa $ and $\protect\lambda %
_{L}=\protect\lambda _{R}=10\protect\kappa $ (dotted line) and $15\protect%
\kappa $(solid line).} \label{Fig. 2}
\end{figure}
When the detector $D_{b}^{S}$ is clicked at time $t_{4}$, $\left\vert \Phi
\right\rangle _{3}$ collapses to
\begin{eqnarray}
\left\vert \Phi \right\rangle _{4} &=&\frac{1}{\Omega ^{4}}\left( \sin
^{2}\left( 2\theta \right) \lambda _{L}^{4}\left\vert 0\right\rangle
_{A}\left\vert 0\right\rangle _{B}\right. +\sin ^{2}\left( 2\theta \right)
\lambda _{R}^{4}\left\vert 2\right\rangle _{A}\left\vert 2\right\rangle _{B}
\notag \\
&&+\left. 2\cos ^{2}\left( 2\theta \right) \lambda _{L}^{2}\lambda
_{R}^{2}\left\vert 1\right\rangle _{A}\left\vert 1\right\rangle _{B}\right) .
\end{eqnarray}%
Note that all the states $\left\vert \Phi \right\rangle _{m},m=1,2,3,4,$ are
not normalized. It is obvious that if $\lambda _{L}=\lambda _{R}$ and $\tan
^{2}(2\theta )=2$, one can obtain the desired state $\left\vert \Psi
\right\rangle _{AB}$. In this case, the probability of getting $\left\vert
\Psi \right\rangle _{AB}$ from $\left\vert \Phi \right\rangle $ (eq. (9))
can be given by
\begin{equation}
P_{3}=\frac{2\left[ \sin ^{4}(2\theta )+2\cos ^{4}(2\theta )\right] }{16}=%
\frac{1}{12}.
\end{equation}%
Hence the total probability of getting the desired state is
\begin{equation}
P_{total}=\frac{1}{12}\prod\limits_{j}\left[ \frac{e^{-\Gamma \tau _{j}}\sin
^{2}\left( \Omega _{k}\tau _{j}\right) \Omega ^{2}}{\Omega _{k}^{2}}\right] .
\end{equation}%
In fact, it is worth noting that the scheme is deterministic because
one can obtain the entangled state once he detects four photons.
$P_{total}$  only corresponds to probability with which one can
detect four photons. In order to intuitionally show the relation
between $P_{total}$ with the time $\tau _{j}$, we perform a
numerical simulation with the following choice of parameters: $\tau
_{j}=\tau $ for all $j$, $\lambda _{L}=\lambda _{R}=10\kappa $ (and
$15\kappa $), $\gamma _{l}=\gamma _{r}=0.1\kappa $. See FIG. 2. One
can find the maximal probability is at the time $\tau =\arctan
\left( \frac{2\Omega _{k}}{\Gamma }\right) /\Omega _{k}$. Our
efficiency is close to those in other preparations of entanglement
of qubits [31,32]. In fact, both the increasement of $\lambda_i$
properly and the improvement of the rate $\kappa/\gamma$ can improve
the total probability, as can also be seen from FIG. 2.

The entangled state of a pair of qutrits is relatively stable
against spontaneous emission because quantum information is encoded
in the degenerate ground states which are not sensitive to
spontaneous emission effect.  In other words, once the entangled
state is prepared, it has longer life.  What is more, one has to
require that the rotated angle $\theta$ of $FS-PBS$ satisfies
$\tan^2(2\theta)=2$ and $\lambda _{L}=\lambda _{R}$ in order to
obtain the maximally entangled qutrits, otherwise the fidelity will
be reduced. In fact, if $\tan^2(2\theta)=2.5$, one can find that the
fidelity of the final state $\left\vert\Psi(\theta)\right\rangle$ is
$F=\left\vert\left\langle\Psi(\theta)\vert\Psi\right\rangle_{AB}\right\vert^2\simeq
0.99$; If $\frac{\lambda _{L}}{\lambda _{R}}=1.1$, the fidelity $F$
of the final state $\left\vert \Psi ^{\prime }\right\rangle $ is
$F=\left\vert \left\langle \Psi ^{\prime }|\Psi \right\rangle
_{AB}\right\vert ^{2}\simeq 0.98$. Both show slight influences. What
is more, the inefficient detections leading to less clicks of the
detectors only reduce the success probability instead of fidelity,
so does the failure of initialization of the initial states of atoms
and cavities. As mentioned in Ref. [25], because the photons from
spontaneous emissions to free modes run with random directions, they
can not be registered by the detectors. Thus the fidelity is not
influenced too. The reduction of the success probability has been
included in our result.

In fact, one can find that the same conclusion can be drawn if one
takes the coincidence detections. It is also worthy of being noted
that if Alice and Bob trap the atoms at their hands in a single
cavity, respectively, the same result can also be obtained so long
as the detections of photons can not provide any information on
which atom a photon is emitted from. In this way, the entangled two
qutrits are trapped in a single cavity respectively which might be
not only more convenient to the further operations of a single
qutrit in the quantum information processing later but it is not
necessary to control 4 cavities in practice. In this way, one can
also avoid the simultaneous preparation four identical cavites+atoms
in practice. Of course, if the four (or two) systems of cavity+atom
are not identical, one has to analyze the fidelity of the final
state.

What we used also consists of linear optical elements, and photon
detectors, which has been widely used to entangle photons. In
particular, the similar optical setups has been used to successfully
prepare W states of photons in experiment [35]. Therefore, our
schemes are feasible by current technologies.

In summary, we have presented a simple scheme to entangle two distant
qutrits only using linear optical elements. The key is the
indistinguishability of photons emitted from the entangled atoms. Since the
quantum information is encoded on the degenerate ground states of a pair of
atoms, the entanglement of the two qutrits is relatively stable. Our scheme
has been shown to be suitable for both cases with and without coincidence
detections. However, the latter can dramatically relax the conditions of the
experimental realization. The scheme allows the two atoms at Alice's or
Bob's side trapped in a single cavity which might be convenient to further
applications. It has been shown that the fidelity is independent of the
inefficient detections, spontaneous emissions and so on.

This work was supported by the National Natural Science Foundation
of China, under Grant No. 10747112, No. 10575017 and No. 60578014.

\end{document}